\documentclass[twocolumn,pra,showpacs,aps,superscriptaddress]{revtex4}
\pdfoutput=1
\usepackage[T1]{fontenc}
\usepackage[latin9]{inputenc}
\usepackage{amsmath}
\usepackage{amssymb}
\usepackage{graphicx}
\usepackage{esint}

\begin{document}

\title{Two-body anticorrelation in a harmonically trapped ideal Bose gas}

\author{T.~M. Wright}
\affiliation{The University of Queensland, School of Mathematics and Physics, Brisbane, Queensland 4072, Australia} 

\author{A. Perrin}
\affiliation{CNRS, UMR 7538, F-93430, Villetaneuse, France}
\affiliation{Universit\'e Paris 13, Sorbonne Paris Cit\'e, Laboratoire de Physique des Lasers, F-93430, Villetaneuse, France}

\author{A. Bray}
\affiliation{The University of Queensland, School of Mathematics and Physics, Brisbane, Queensland 4072, Australia} 

\author{J. Schmiedmayer}
\affiliation{Vienna Center for Quantum Science and Technology, Atominstitut, TU Wien, 1020 Vienna, Austria} 

\author{K.~V. Kheruntsyan}
\affiliation{The University of Queensland, School of Mathematics and Physics, Brisbane, Queensland 4072, Australia}

\begin{abstract}
We predict the existence of a dip below unity in the second-order coherence function of a partially condensed ideal Bose gas in harmonic confinement, signaling the anticorrelation of density fluctuations in the sample.  The dip in the second-order coherence function is revealed in a canonical-ensemble calculation, corresponding to a system with fixed total number of particles.   In a grand-canonical ensemble description, this dip is obscured by the occupation-number fluctuation catastrophe of the ideal Bose gas.  The anticorrelation is most pronounced in highly anisotropic trap geometries containing small particle numbers.  We explain the fundamental physical mechanism which underlies this phenomenon, and its relevance to experiments on interacting Bose gases. 
\end{abstract}

\pacs{05.30.Jp, 03.75.Hh}

\date{\today}

\maketitle

\section{Introduction}

The ``grand-canonical fluctuation catastrophe'' \cite{Holthaus:98} of the ideal Bose gas has been the subject of longstanding attention since the early days of quantum statistical mechanics \cite{Einstein2a,Uhlenbeck,Kahn-Uhlenbeck,Fierz:56,Fujiwara:70,Ziff} (for a comprehensive review see Ref.~\cite{Scully:06}).  The epithet \emph{catastrophe} refers to the pathologically large occupation-number fluctuations predicted for the condensate mode below the condensation temperature $T_C$ in the grand canonical ensemble.  Such a prediction is pathological because the fluctuations must in fact become small and eventually vanish as atoms condense into the ground state as the temperature $T$ approaches zero.  This anomaly is resolved by the introduction of weak interparticle interactions, but for a strictly ideal Bose gas the problem must be considered within the canonical or microcanonical formalisms \cite{Fierz:56,Fujiwara:70,Ziff,Politzer:96,Wilkens1,Wilkens2,Holthaus:98,Scully:06}, corresponding to a diffusively isolated system with a fixed total number of particles, in order to obtain physically sensible results. 

The advent of Bose-Einstein condensates (BECs) in trapped atomic gas experiments~\cite{JILA-BEC,Rice-BEC,MIT-BEC} has lead to a revival of interest in the condensate number fluctuation problem~\cite{Politzer:96,Grossmann:96,Wilkens1,Wilkens2,Gajda:97,Holthaus:98,Scully:06}, as the isolation of these systems by the trapping potential makes them practically ideal realizations of the restricted statistical ensembles with finite and well-defined total particle numbers. Parallel developments of experimental techniques to probe these systems via the measurement of atom-field correlations~\cite{Shimizu:96,Hannover:03,Schellekens:05,g2-Esslinger:05,Jeltes:07,Bloch-fermi-antibunching,Manz:10,Dall:10,Hodgman:11,Perrin-HBT:12,g2-Tolra:04,g2-Weiss:05,g2-Greiner:05,g2-Bloch:05,Perrin:07,Vienna-twins} such as the two-point, two-body coherence function~\cite{Nar-Glauber:99}  
\begin{equation}\label{eq:g2r-def} 
    g^{(2)}(\mathbf{r},\mathbf{r}^{\prime})=\frac{\langle\hat{\Psi}^{\dagger}(\mathbf{r})\hat{\Psi}^{\dagger}(\mathbf{r}^{\prime})\hat{\Psi}(\mathbf{r^{\prime}})\hat{\Psi}(\mathbf{r})\rangle}{\langle\hat{\Psi}^{\dagger}(\mathbf{r})\hat{\Psi}(\mathbf{r})\rangle\langle\hat{\Psi}^{\dagger}(\mathbf{r^{\prime}})\hat{\Psi}(\mathbf{r}^{\prime})\rangle},
\end{equation}
where $\hat{\Psi}^{\dagger}(\mathbf{r})$ and $\hat{\Psi}(\mathbf{r})$ are the bosonic creation and annihilation field operators, respectively, allow us to reformulate the problem in terms of the spatial coherence of the Bose field. 

\vspace{\baselineskip}
Within the grand-canonical description of the ideal Bose gas one finds that the second-order coherence at zero spatial separation is given by $g^{(2)}(\mathbf{r},\mathbf{r})=2$~\cite{Nar-Glauber:99,Gomes:06}, even in the limit $T\to 0$. This is simply the matter-wave analog of the celebrated Hanbury Brown and Twiss (HBT) effect of photon ``bunching''~\cite{HBT:56,HBT-stellar:56} --- a two-particle interference effect due to the fundamental symmetry of many-body Bose wavefunctions~\cite{Baym98}.  This prediction, while being correct for high-temperature gases~\cite{Schellekens:05,g2-Esslinger:05,Dall:10,Perrin-HBT:12}, is at variance with the expectation that a Bose system at temperatures well below $T_C$ should exhibit higher-order coherence and thus $g^{(2)}(\mathbf{r},\mathbf{r})\approx1$ (see, e.g., Refs.~\cite{Burt97,Ketterle97a}).  This discrepancy arises directly from the spuriously large fluctuations of the condensate occupation in the grand-canonical ensemble, and is therefore resolved in a noninteracting gas by adopting a canonical-ensemble treatment of the atom-number fluctuations in the ground state~\cite{Nar-Glauber:99}, or by the introduction of weak interparticle interactions.   

In this article we show that the canonical ensemble treatment of the ground state (condensate mode) occupation fluctuations also reveals an additional and counterintuitive effect: an \emph{anticorrelation} dip $g^{(2)}(\mathbf{r},\mathbf{r}^{\prime})<1$ in the two-body coherence function evaluated at particular finitely separated ($|\mathbf{r}-\mathbf{r}^{\prime}|>0$) points $\mathbf{r}$ and $\mathbf{r}'$. This is a somewhat surprising result for noninteracting, bosonic atoms, as density anticorrelation effects are usually associated with either strong repulsive interactions in bosonic systems~\cite{Sykes:08} or Pauli blocking in fermionic systems~\cite{Bloch-fermi-antibunching}.  We show here that the anticorrelation of density fluctuations in the trapped ideal Bose gas arises from the intrinsic \emph{amplitude} anticorrelation of excited single-particle trap eigenstates through their interference with the condensate mode.  This effect becomes pronounced when the lowest lying excited trap eigenstates are highly populated, but requires also that the condensate mode itself exhibits partial second-order coherence, and is thus obscured by the fluctuation catastrophe in a grand-canonical calculation of $g^{(2)}(\mathbf{r},\mathbf{r}')$. 

The two-body coherence function of a harmonically trapped ideal Bose gas has previously been analyzed in Refs.~\cite{Nar-Glauber:99,Gomes:06}.  In these works, the canonical ensemble treatment of the condensate-mode population fluctuations was implemented only to the leading order --- an approximation that is well justified for systems containing a very large number ($N\sim10^{6}$) of atoms. While this is sufficient for describing the primary effect of condensation on $g^{(2)}(\mathbf{r},\mathbf{r}')$ --- the establishment of local second-order coherence $g^{(2)}(\mathbf{r},\mathbf{r})\approx1$ --- the restriction to very large $N$ also implies a vanishingly small anticorrelation dip. As we show here, the anticorrelation dip reaches appreciable sizes only in systems containing $\sim 100$--$1000$ atoms.  In this regime, however, the leading-order approximation of the condensate population fluctuations becomes inadequate for quantitatively predicting the magnitude of the dip.  Here we treat the condensate-mode fluctuations exactly, using a numerical implementation of a recurrence relation for the canonical partition function~\cite{Landsberg:61,Brosens:96,Borrmann:97,Wilkens2} and its relation to the full counting statistics of the condensate mode population~\cite{Wilkens1,Wilkens2,Scully:06}. Our treatment therefore 
provides a more complete and quantitatively accurate description of the two-body coherence function of a harmonically trapped ideal Bose gas.  As we discuss in Sec.~\ref{sec:Relation-to-practice}, the ideal-gas regime in which a canonical treatment of fluctuations is necessitated, and in which the anticorrelation dip is expected to occur, is within reach of modern experimental atomic-physics techniques.  Moreover, the ideal-gas analysis we present here serves as an instructive prototype for closely related anticorrelation effects in interacting-gas experiments~\cite{Perrin-HBT:12}, which will be discussed elsewhere~\cite{interacting_case}.

\section{Two-body correlation of an ideal Bose gas in a harmonic trap}

We focus here on a system of $N$ noninteracting Bose atoms confined to a three-dimensional (3D) harmonic trapping potential with frequencies $\omega_{i}\;(i=x,y,z)$.  We denote the eigenstates of the trap by $\zeta_n(\mathbf{r})\, (n\geq0)$, and their corresponding energies by $E_n$.  The annihilation operator $\hat{a}_n=\int\!d\mathbf{r}\,\zeta_n^*(\mathbf{r})\hat{\Psi}(\mathbf{r})$ destroys an atom in the state $\zeta_n(\mathbf{r})$.  In Appendix~\ref{append:full_g2} we derive an approximate form 
\begin{equation}\label{eq:g2-IBG} 
    g^{(2)}(\mathbf{r},\mathbf{r}^{\prime})=1+\frac{\left|G^{(1)}(\mathbf{r},\mathbf{r}^{\prime})\right|^{2}}{\rho(\mathbf{r})\rho(\mathbf{r}^{\prime})}-\left(2-g_{0}^{(2)}\right)\frac{\rho_{0}(\mathbf{r})\rho_{0}(\mathbf{r}^{\prime})}{\rho(\mathbf{r})\rho(\mathbf{r}^{\prime})},
\end{equation}
for the second-order coherence function of the gas, neglecting higher-order terms that have little effect on the anticorrelation dip we study in this article.  Here, $G^{(1)}(\mathbf{r},\mathbf{r}') \equiv \langle \hat{\Psi}^\dagger(\mathbf{r})\hat{\Psi}(\mathbf{r}')\rangle$ is the first-order coherence function of the gas, $\rho(\mathbf{r})=G^{(1)}(\mathbf{r},\mathbf{r})$ is the atomic density, $\rho_0(\mathbf{r})=\langle \hat{a}_0^\dagger \hat{a}_0\rangle |\zeta_0(\mathbf{r})|^2$ is the density of atoms in the ground state, and $g_n^{(2)} \equiv \langle \hat{a}_n^\dagger \hat{a}_n^\dagger \hat{a}_n\hat{a}_n\rangle/\langle\hat{a}_n^\dagger \hat{a}_n\rangle^2$ is the second-order coherence of the mode $\zeta_n(\mathbf{r})$.   

All expectation values in Eq.~(\ref{eq:g2-IBG}) refer, in principle, to traces with respect to the density matrix appropriate to the canonical ensemble.  Our analysis is, however, substantially simplified by calculating all \emph{occupation numbers} in the grand-canonical ensemble with mean total atom number $N$.  Although the grand-canonical ensemble greatly overestimates the fluctuations of the condensate population, it predicts the mean occupations of the trap modes with very little error, compared to the exact canonical-ensemble values (see, e.g., Refs.~\cite{Politzer:96,Wilkens1}).  Following Ref.~\cite{Ketterle:96}, for a given atom number $N$ and temperature $T$, we calculate occupation numbers in the grand-canonical ensemble with chemical potential $\mu$ determined by the implicit equation 
\begin{equation}
    N=\int\!\rho(\mathbf{r})\,d\mathbf{r}=\sum_{l=1}^{\infty}e^{l\mu/k_{B}T}\prod_{i}\frac{1}{\left(1-e^{-l\epsilon_{i}}\right)},
\end{equation}
where we have introduced the notation $\epsilon_{i}=\hbar\omega_{i}/k_{B}T$ for compactness.  We can therefore use the known grand-canonical expressions for the first-order coherence function~\cite{Landau-Lifshitz-StatPhys,Nar-Glauber:99}  
\begin{gather}\label{eq:G1} 
    G^{(1)}(\mathbf{r},\mathbf{r}^{\prime})=\sum_{l=1}^{\infty}e^{l\mu/k_{B}T}\nonumber \\
    \times\prod_{i}\frac{\exp\left[-\tanh\left(\frac{l\epsilon_{i}}{2}\right)\left(\frac{r_{i}+r_{i}^{\prime}}{2\sigma_{i}}\right)^{2}-\coth\left(\frac{l\epsilon_{i}}{2}\right)\left(\frac{r_{i}-r_{i}^{\prime}}{2\sigma_{i}}\right)^{2}\right]}{\sqrt{\pi}\sigma_{i}\sqrt{1-e^{-2l\epsilon_{i}}}},
\end{gather}
and the atomic density
\begin{equation}\label{eq:rho} 
    \rho(\mathbf{r})=\sum_{l=1}^{\infty}e^{l\mu/k_{B}T}\prod_{i}\frac{\exp\left[-\tanh\left(\frac{l\epsilon_{i}}{2}\right)\frac{r_{i}^{2}}{\sigma_{i}^{2}}\right]}{\sqrt{\pi}\sigma_{i}\sqrt{1-e^{-2l\epsilon_{i}}}},
\end{equation}
where $\sigma_{i}=\sqrt{\hbar/m\omega_{i}}$ is the characteristic length of the harmonic-oscillator potential in the $i$th direction.  Moreover, the ground-state (condensate) occupation is given in this approach by $N_{0}=\langle\hat{a}_{0}^{\dagger}\hat{a}_{0}\rangle=(e^{-\mu/k_{B}T}-1)^{-1}$, and thus the ground state density 
\begin{equation}\label{eq:rho0} 
    \rho_{0}(\mathbf{r})=\frac{1}{e^{-\mu/k_{B}T}-1}\prod_{i}\frac{e^{-r_{i}^{2}/\sigma_{i}^{2}}}{\sqrt{\pi}\sigma_{i}}.
\end{equation}

Implicit in Eq.~(\ref{eq:g2-IBG}) is the assumption that the number fluctuations of excited modes are well-approximated by the usual grand-canonical predictions, and in particular that no correlation exists between the fluctuations of any two distinct modes.  The validity of this assumption for the purposes of exhibiting the anticorrelation dip is discussed in Appendix~\ref{append:full_g2}.  Equation~(\ref{eq:g2-IBG}) does, however, explicitly account for the canonical-ensemble number fluctuations of the ground state, which we quantify by~\cite{Walls:08}   
\begin{equation}\label{eq:g2-condensate} 
    g_{0}^{(2)}=\frac{\langle\hat{a}_{0}^{\dagger}\hat{a}_{0}^{\dagger}\hat{a}_{0}\hat{a}_{0}\rangle}{\langle\hat{a}_{0}^{\dagger}\hat{a}_{0}\rangle^{2}}=1-\frac{1}{N_{0}}+\frac{(\triangle N_{0})^{2}}{N_{0}^{2}},
\end{equation}
where $\Delta N_{0}=[\langle N_0^2 \rangle - \langle N_0 \rangle^2]^{1/2}$ is the root-mean-square (rms) dispersion of $N_0$.  

Neglecting the last term on the right-hand side of Eq.~(\ref{eq:g2-IBG}) we regain the usual grand-canonical form for $g^{(2)}(\mathbf{r},\mathbf{r}')$, from which we find, for any $T>0$, a monotonically decreasing profile of the two-body coherence $g^{(2)}(\mathbf{r},\mathbf{r}+\Delta\mathbf{r})$ as a function of the relative separation $\Delta\mathbf{r}=\mathbf{r}^{\prime}-\mathbf{r}$, with the coherence function decreasing from its peak value of $g^{(2)}(\mathbf{r},\mathbf{r})=2$ at zero separation to $g^{(2)}(\mathbf{r},\mathbf{r}+\Delta\mathbf{r})=1$ at infinitely large separations $|\Delta \mathbf{r}|$.  While this behavior is valid above the transition temperature $T_C$, the prediction that $g^{(2)}(\mathbf{r},\mathbf{r})=2$ even at temperatures $T<T_C$ is in contrast to the partially coherent behavior $g^{(2)}(\mathbf{r},\mathbf{r})\lesssim2$ expected in the presence of a Bose condensate~\cite{Baym98} --- a manifestation of the grand-canonical fluctuation catastrophe in terms of the two-body coherence function. 

\begin{figure}[tb]
    \includegraphics[width=8.2cm]{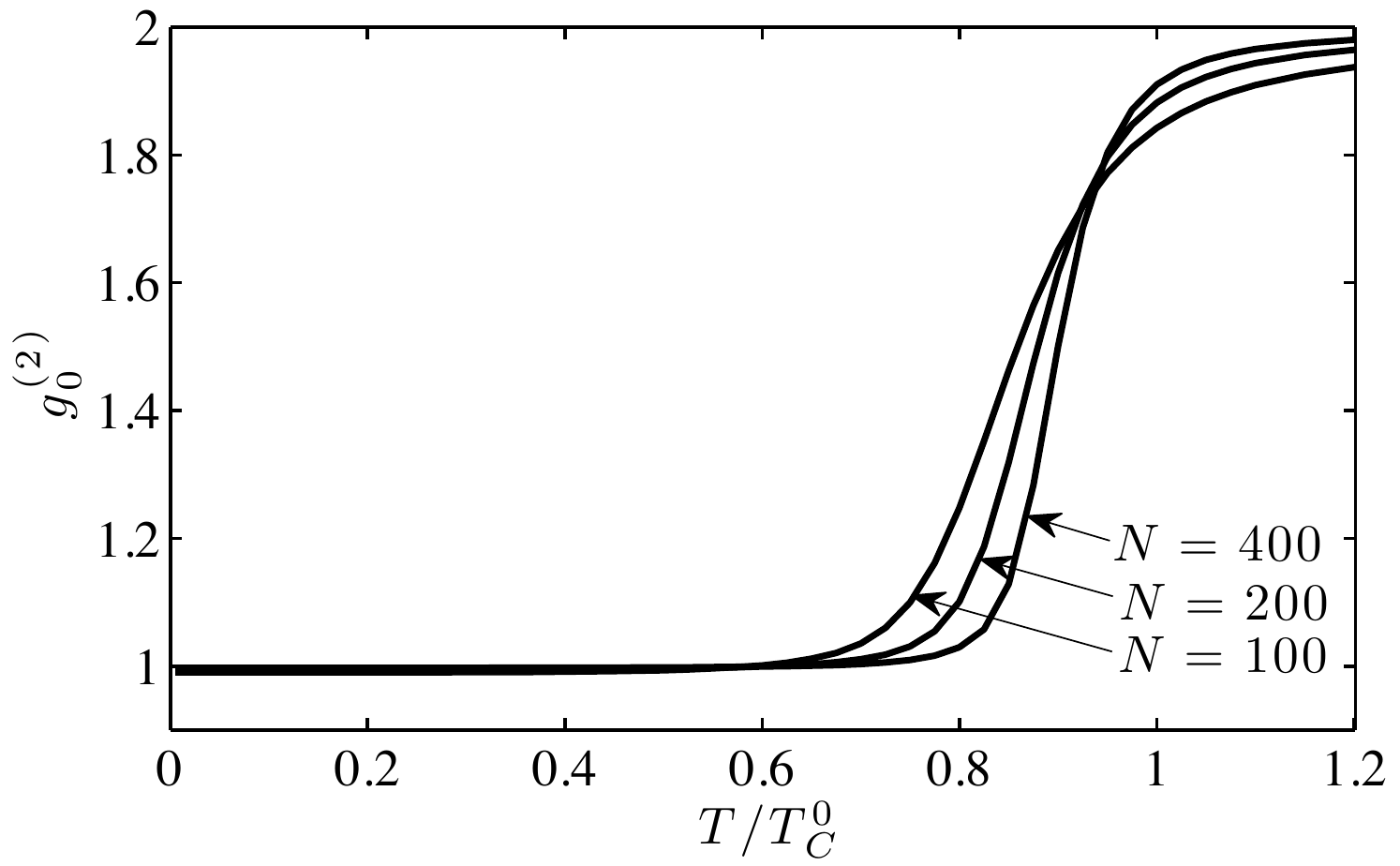}
    \caption{Second-order coherence $g_{0}^{(2)}$ [Eq.~(\ref{eq:g2-condensate})] of the condensate mode in an isotropic 3D harmonic trapping potential ($\omega_i\equiv\bar{\omega},\,i=x,y,z$), as a function of the reduced temperature $T/T_C^0$, where $T_C^0\equiv T_C^0(N)$ is the thermodynamic-limit value for the ideal-gas condensation temperature [Eq.~(\ref{eq:Tc0})].  The three curves correspond to three distinct values of the total atom number $N$.} 
    \label{fig:1} 
\end{figure}

The approximate form~(\ref{eq:g2-IBG}) of the second-order coherence function generalizes the one employed in Refs.~\cite{Nar-Glauber:99,Gomes:06} that itself includes a leading-order description of the suppression of condensate number fluctuations below the critical temperature.  Specifically, the form of $g^{(2)}(\mathbf{r},\mathbf{r}')$ used in those references is obtained from Eq.~(\ref{eq:g2-IBG}) in the limit that the condensate mode exhibits Poissonian number fluctuations (i.e., that $g^{(2)}_0=1$).  Such an approach neglects the fact that the fluctuations of the condensate occupation vary smoothly (in a finite system) from those of a thermal mode ($g^{(2)}_0=2$) to Poissonian as the system temperature is lowered through the Bose-condensation transition.  The regime of intermediate fluctuation statistics $1<g^{(2)}_0<2$ becomes increasingly narrow as the total atom number $N$ increases, and approaches a sudden step-change at the critical temperature in the thermodynamic limit.  The approximation $g^{(2)}_0=1$ is thus well justified for large atom numbers, as the second-order coherence of the condensate mode deviates significantly from this value only over a very narrow temperature range on the condensed side of the transition, whereas above $T_C$ (where $g_0^{(2)}\approx 2$) the spurious contribution of the last term in Eq. (\ref{eq:g2-IBG}) is negligible anyway, as $\rho_{0}({\mathbf{r}})\ll\rho({\mathbf{r}})$ in this regime~\cite{NoteA}. 

For smaller total atom numbers, however, the crossover region becomes broad and precise knowledge of $g_{0}^{(2)}$ is required to accurately calculate the two-body correlation function $g^{(2)}(\mathbf{r},\mathbf{r}^{\prime})$.  Accordingly, it is crucial to evaluate the variance $(\Delta N_{0})^{2}$ and hence $g_{0}^{(2)}$ with as few approximations as possible.  Here we employ a recurrence relation for the canonical partition function \cite{Wilkens1,Wilkens2,Scully:06} to numerically calculate the canonical-ensemble values of $(\Delta N_{0})^{2}$ (and thus $g_{0}^{(2)}$) exactly~\cite{NoteB}. 

\begin{figure}[tb]
    \includegraphics[width=8.5cm]{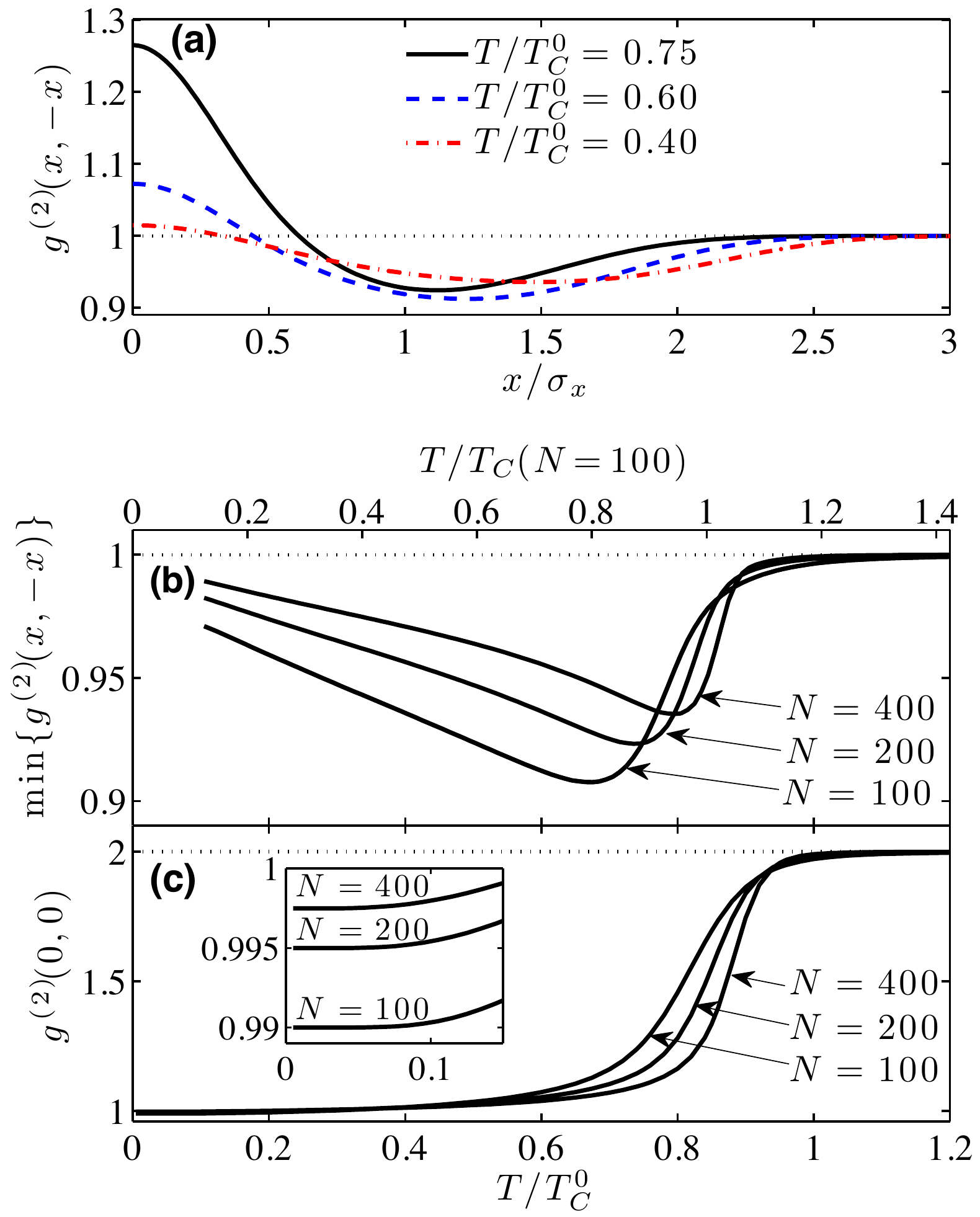}
    \caption{(Color online) (a) Symmetrically evaluated nonlocal coherence function $g^{(2)}(x,-x)\equiv g^{(2)}(x\hat{\mathbf{x}},-x\hat{\mathbf{x}})$ for an ideal Bose gas of $N=100$ atoms in an isotropic 3D harmonic trapping potential ($\omega_i\equiv\bar{\omega},\,i=x,y,z$).  The three curves correspond to three distinct reduced temperatures $T/T_C^0$, where $T_C^0\equiv T_C^0(N)$ is the thermodynamic limit value for the ideal-gas condensation temperature [Eq.~(\ref{eq:Tc0})].  (b) The minimal value of $g^{(2)}(x,-x)$ as a function of $T/T_C^{0}$, for $N=100,\;200$, and $400$ atoms.  (c) The value of the local two-body coherence function at the trap center, $g^{(2)}(0,0)$.  The upper horizontal axis in (b) shows $T$ as a proportion of the finite-size corrected critical temperature $T_C$ [Eq. (\ref{eq:Tc})] for $N=100$ atoms.  The inset shows the behavior of the second-order coherence $g^{(2)}(0,0)$ at the trap center as $T\to 0$, where we find $g^{(2)}(0,0)\to1-1/N$ (see text).}  
    \label{fig:2} 
\end{figure}

\subsection{Isotropic trap}

We consider first the case of an isotropic 3D trapping potential ($\omega_i\equiv\bar{\omega},\,i=x,y,z$), and show in Fig.~\ref{fig:1} the calculated values of $g_{0}^{(2)}$ as functions of the temperature $T$, for $N=100,\;200,$~and~$400$ atoms.  For these small atom numbers the crossover from $g^{(2)}_0\approx 2$ to $g^{(2)}_0\approx 1$ occurs over a broad range of temperatures.  We quote the temperature in Fig.~\ref{fig:1} as a fraction of the thermodynamic-limit result for the transition temperature of the ideal Bose gas with $N$ atoms   
\begin{equation}\label{eq:Tc0}
    T_{C}^{0}=\frac{\hbar}{k_{B}}\left(\frac{N\omega_{x}\omega_{y}\omega_{z}}{\zeta(3)}\right)^{1/3},
\end{equation}
where the Riemann zeta function $\zeta(3)\approx1.202$. The temperature $T_C^0$ corresponds to the onset of saturation of the excited-state population and is easily derived in the semiclassical limit $k_{B}T\gg\hbar\omega_{i}$ (see, e.g., Ref.~\cite{Pethick02}).  In systems with a relatively small $N$, finite-size effects result in a lowering of the effective condensation temperature to~\cite{Ketterle:96}   
\begin{equation}\label{eq:Tc} 
    T_{C}=T_{C}^{0}\left(1-\frac{0.7275\sum_{i}\omega_{i}}{3N^{1/3}(\prod_{i}\omega_{i})^{1/3}}\right),
\end{equation}
and we find $T_C/T_C^0 \approx 0.840,\;0.877$,~and~$0.901$, for $N=100,\;200,$~and~$400$ atoms, respectively.  

We note that the forms of $g^{(2)}_0$ presented in Fig.~\ref{fig:1} are reminiscent of the behavior of condensate fluctuations near the transition to condensation in the \emph{interacting} Bose gas (see Ref.~\cite{Wright11} and references therein).  The fact that such behavior is also obtained for the canonical \emph{ideal} Bose gas is consistent with the arguments of Wilkens and Weiss~\cite{Wilkens1} that the constraint of fixed particle number $N$ in the canonical ensemble acts as an effective interparticle interaction that bestows a second-order character upon the ideal-gas transition~\cite{NoteC}.   

In Fig.~\ref{fig:2}(a) we plot the symmetrically evaluated nonlocal coherence function $g^{(2)}(x,-x)\equiv g^{(2)}(x\hat{\mathbf{x}},-x\hat{\mathbf{x}})$, calculated using Eqs.~(\ref{eq:g2-IBG})--(\ref{eq:g2-condensate}) for the same parameters as the results of Fig.~\ref{fig:1}.  An anticorrelation dip $g^{(2)}(x,-x)<1$ at finite $x$ is clearly visible for all three temperatures considered.  This dip can easily be understood in terms of two-particle interference between excited modes of the trap and the condensate mode.  Any given excited mode of the trap $\zeta_n(\mathbf{r}),\;n>0$ has a spatial structure, varying from positive to negative with $\mathbf{r}$ (for simplicity we consider the real Hermite modes of the trap).  Thus when this mode interferes constructively with the condensate (which has a uniform phase across its extent) at, e.g., a point where $\zeta_n(\mathbf{r})>0$, it must correspondingly interfere \emph{destructively} at a point where $\zeta_n(\mathbf{r})<0$.  This interference produces a positive density fluctuation at the former point, and a negative density fluctuation at the latter point, and provides the basic mechanism for the anticorrelation of density fluctuations between the two points.  However, the condensate mode must be at least partially coherent ($g^{(2)}_0<2$) in order for this tendency towards anticorrelation to overcome the (positive) correlation between fluctuations at the two points due to the HBT effect.  Thus the anticorrelation dip is not observed in a grand-canonical treatment, due to the fluctuation catastrophe.  A more detailed discussion of this fundamental mechanism is given in Appendix~\ref{append:physical_origin}. 

In Fig.~\ref{fig:2}(b) we plot the minimum value of $g^{(2)}(x,-x)$ as a function of the temperature of the gas.  For comparison, the second-order coherence $g^{(2)}(0,0)$ at the trap center is shown in Fig.~\ref{fig:2}(c).  By the arguments in Appendix~\ref{append:physical_origin} we expect that the magnitude of the dip in $g^{(2)}(x,-x)$ below the uncorrelated level of unity is, at leading order, approximately proportional to the peak fractional occupation $N_1/N$ of the first excited state (along $x$), which can be estimated to scale roughly as $O(1/\sqrt{N})$~\cite{Ketterle:96}, and the results of Fig.~\ref{fig:2}(b) are consistent with this scaling.  The anticorrelation is thus more pronounced for systems containing smaller numbers of atoms, and becomes essentially negligible for $N\sim10^{6}$ atoms.  We also note that the second-order coherence $g^{(2)}(0,0)$ at the trap center does not precisely approach unity as $T\to0$ [inset to Fig.~\ref{fig:2}(c)], but in fact tends towards the $g^{(2)}(\mathbf{r},\mathbf{r})=1-1/N$ result appropriate to a Fock state of $N$ atoms~\cite{Walls:08}, illustrating the precision to which we obtain the second-order coherence in this limit. 

A second dependence of the anticorrelation dip on the total atom number $N$ is that the peak anticorrelation effect occurs at progressively higher temperatures as the atom number is increased.  This can also be understood on the basis that the magnitude of the anticorrelation dip should be approximately proportional to $N_1$:  The peak value of $N_1$ in the ideal gas occurs at increasingly higher temperatures approaching $T_C^0$ as the atom number $N\to\infty$~\cite{Ketterle:96}, and indeed the temperature at which the maximal anticorrelation appears is reasonably well predicted by the temperature at which the product $N_0N_1$ (not shown) attains its peak value, though is somewhat skewed to lower temperatures due to the dependence of $g_0^{(2)}$ on $T$.  We note also that the dip of $g^{(2)}(x,-x)$ below unity broadens and becomes centered around increasing $x$ as the temperature of the system is lowered.  This is easily understood in terms of decreasing (positive) contributions to $g^{(2)}(x,-x)$ due to HBT bunching of atoms in excited modes, which obscure the anticorrelation effect, as the temperature is reduced. 

\begin{figure}[tb]
    \includegraphics[width=8.5cm]{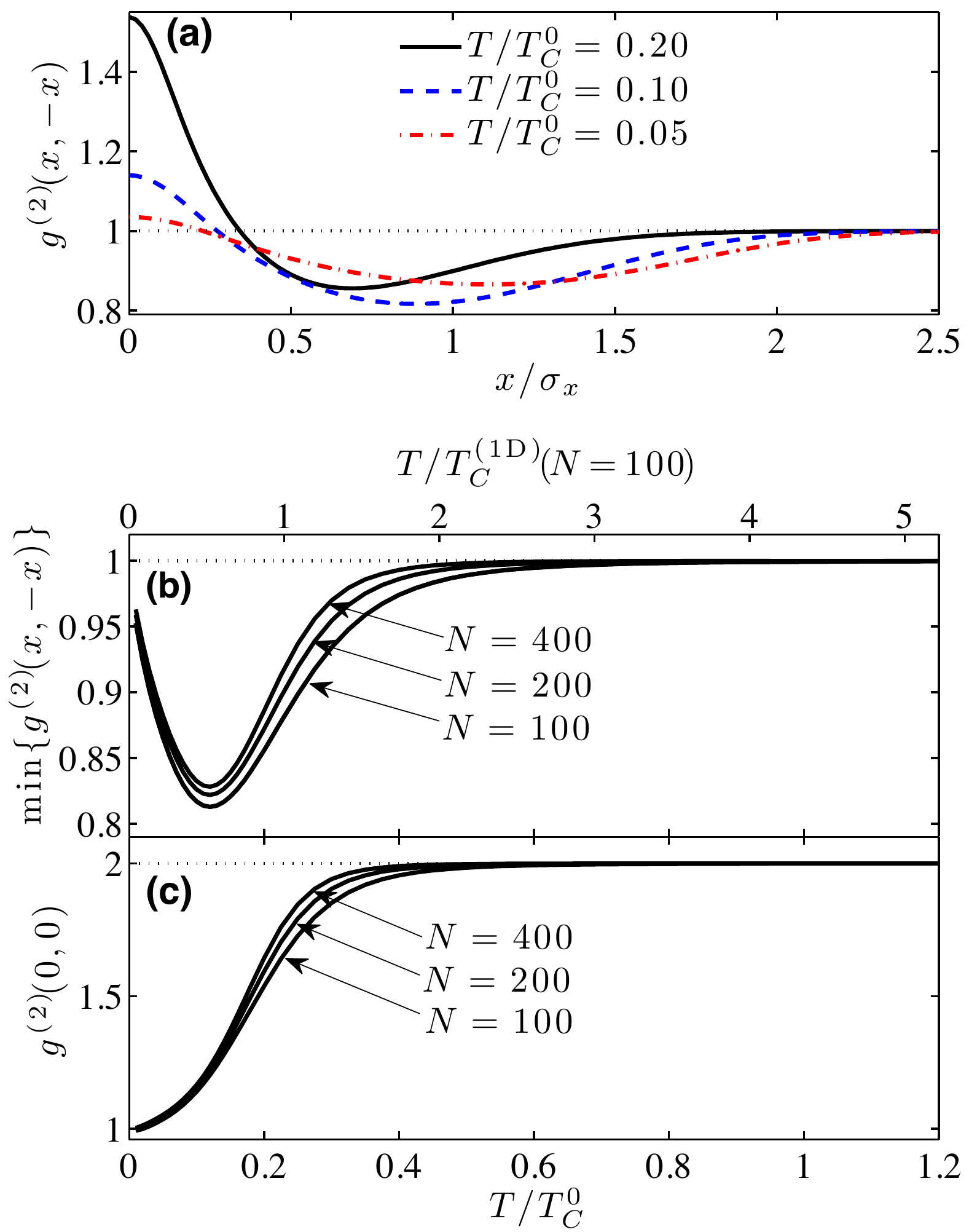}
    \caption{(Color online) (a) Symmetrically evaluated nonlocal coherence function $g^{(2)}(x,-x)\equiv g^{(2)}(x\hat{\mathbf{x}},-x\hat{\mathbf{x}})$ for an ideal Bose gas of $N=100$ atoms in a highly anisotropic trapping potential ($\omega_\perp/\omega_x=127$).  The three curves correspond to three distinct reduced temperatures $T/T_C^0$.  (b) The minimal value of $g^{(2)}(x,-x)$ as a function of $T/T_C^0$, for $N=100,\;200,$~and~$400$ atoms, in traps with aspect ratios $\omega_\perp/\omega_x=127,\;223,$~and~$395$, respectively.  (c) The value of the two-body correlation function at the trap center $g^{(2)}(0,0)$ as a function of $T/T_{C}^{0}$, for $N=100$, $200$, and $400$ atoms. The upper horizontal axis in (b) shows $T$ in units of the 1D critical temperature $T_{C}^{\mathrm{(1D)}}$ for the case of $N=100$ atoms.}
    \label{fig:3} 
\end{figure}

\subsection{Highly anisotropic trap and 1D systems}\label{subsec:quasi-1D}

We now analyze the two-body correlation function of an ideal Bose gas in a highly anisotropic (cigar-shaped) trapping potential. This is a particularly important case as the anisotropic confinement, for temperatures well below the excitation energy in the tightly confined transverse dimensions ($k_{B}T\ll\hbar\omega_{\bot}$, where we have assumed $\omega_{y}=\omega_{z}\equiv\omega_{\bot}\gg\omega_{x}$), gives access to one-dimensional (1D) physics which can be quite distinct from 3D physics~\cite{g2-Tolra:04,Bloch-Tonks,g2-Weiss:05,Amerongen,Armijo:11,Jacqmin:11}.  For example, in such a system in the ideal gas regime, provided that $N[\zeta(3)]^{1/2}[\ln(2N)]^{-3/2}<\omega_{\bot}/\omega_{x}$, the role of the 3D critical transition temperature $T_{C}$ (which can be approximated to leading order by $T_{C}^{0}$) is reduced to signaling the onset of the so-called transverse condensation~\cite{Ketterle:97}.  In this regime, atoms accumulate in the transverse ground state below $T_C$ due to the saturation of population in the transverse excited states, yet the system remains noncondensed with respect to the longitudinal states until the effective 1D condensation temperature $T_{C}^{\mathrm{(1D)}}\approx N\hbar\omega_{x}/[k_{B}\ln(2N)]$~\cite{Ketterle:96,Ketterle:97} is reached.  

\begin{figure}[tb]
    \includegraphics[width=8.2cm]{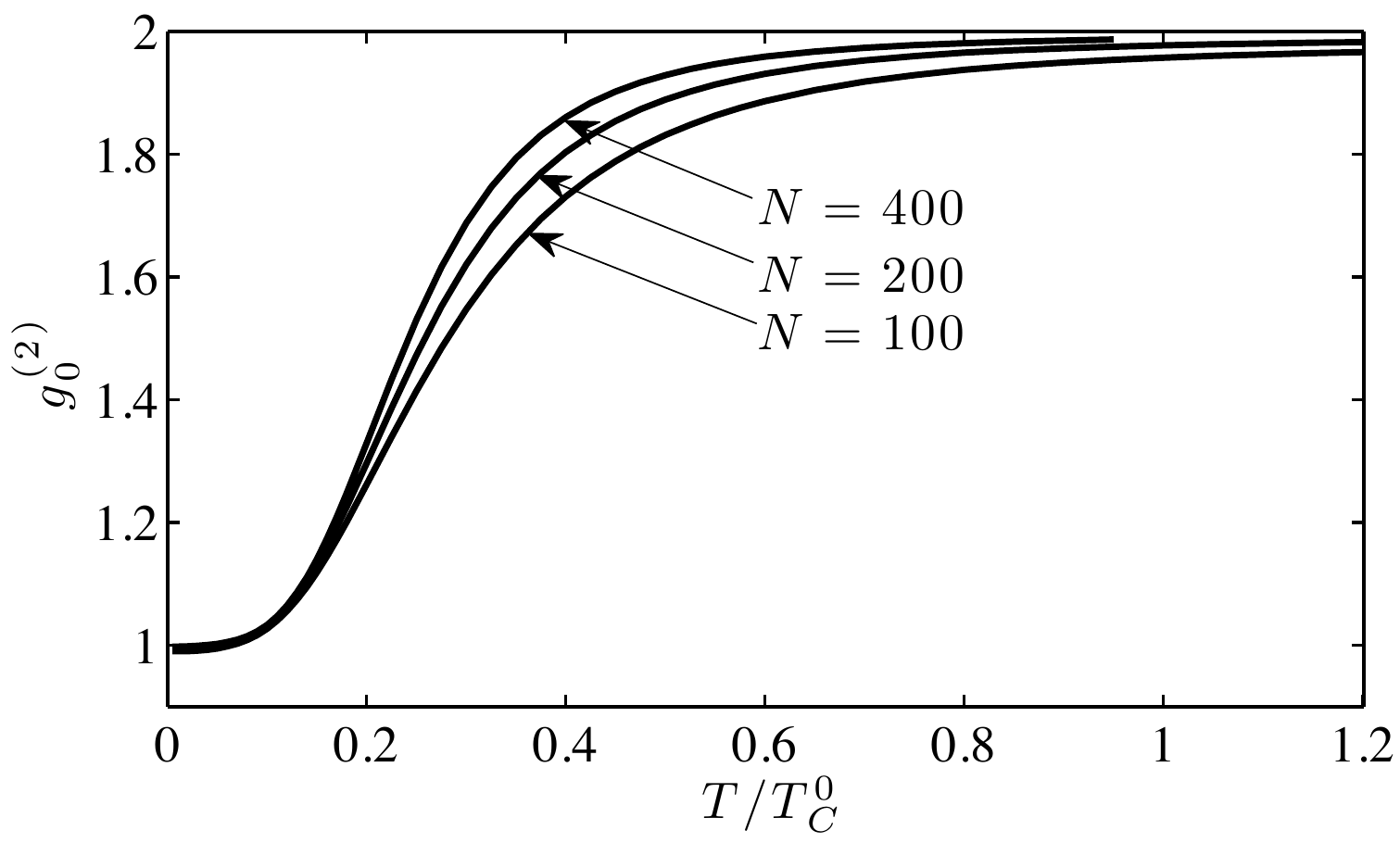}
    \caption{Second-order coherence $g_{0}^{(2)}$ of the condensate mode [Eq.~(\ref{eq:g2-condensate})], as a function of $T/T_{C}^{0}$, for the same parameters as in Fig.~\ref{fig:3}.} 
    \label{fig:4} 
\end{figure}

Examples of the two-body correlation function $g^{(2)}(x,-x)$ symmetrically evaluated along the long axis of a highly anisotropic trap are shown in Fig.~\ref{fig:3}(a).  In all three cases shown we choose $\omega_\perp$ to maintain the condition $k_{B}T_{C}^{\mathrm{(1D)}}/\hbar\omega_{\bot}=0.2$, so that the system is in the quasi-1D regime at temperatures near $T_{C}^{\mathrm{(1D)}}$ and below (we find $T_C^{\mathrm{(1D)}}/T_C^0 \approx 0.230,\;0.220,$~and~$0.212$ for $N=100,\;200,$~and~$400$, respectively). 

We observe [Fig.~\ref{fig:3}(b)] that the strongest anticorrelation occurs near $T\sim T_{C}^{\mathrm{(1D)}}$, in contrast to the isotropic case in which it occurs near $T\sim T_{C}$.  For the same total number of atoms $N$, the maximal magnitude of the anticorrelation dip is larger in the highly anisotropic trap than in a spherically symmetric trap, and as in the isotropic case the dip is deeper for smaller $N$.  In Fig.~\ref{fig:4} we show the values of the second-order coherence function $g^{(2)}_0$ used to calculate the results of Fig.~\ref{fig:3}.  We note that the crossover from $g_{0}^{(2)}\approx2$ to $g_{0}^{(2)}\approx1$ occurs over a very broad range of temperatures, indicating the importance of using precise values for $g^{(2)}_0$ to accurately evaluate $g^{(2)}(\mathbf{r},\mathbf{r}^{\prime})$ in this quasi-1D regime.  The resulting behavior of the second-order coherence $g^{(2)}(0,0)$ at the trap center is shown in Fig.~\ref{fig:3}(c). 

\begin{figure}[tb]
    \includegraphics[width=8.3cm]{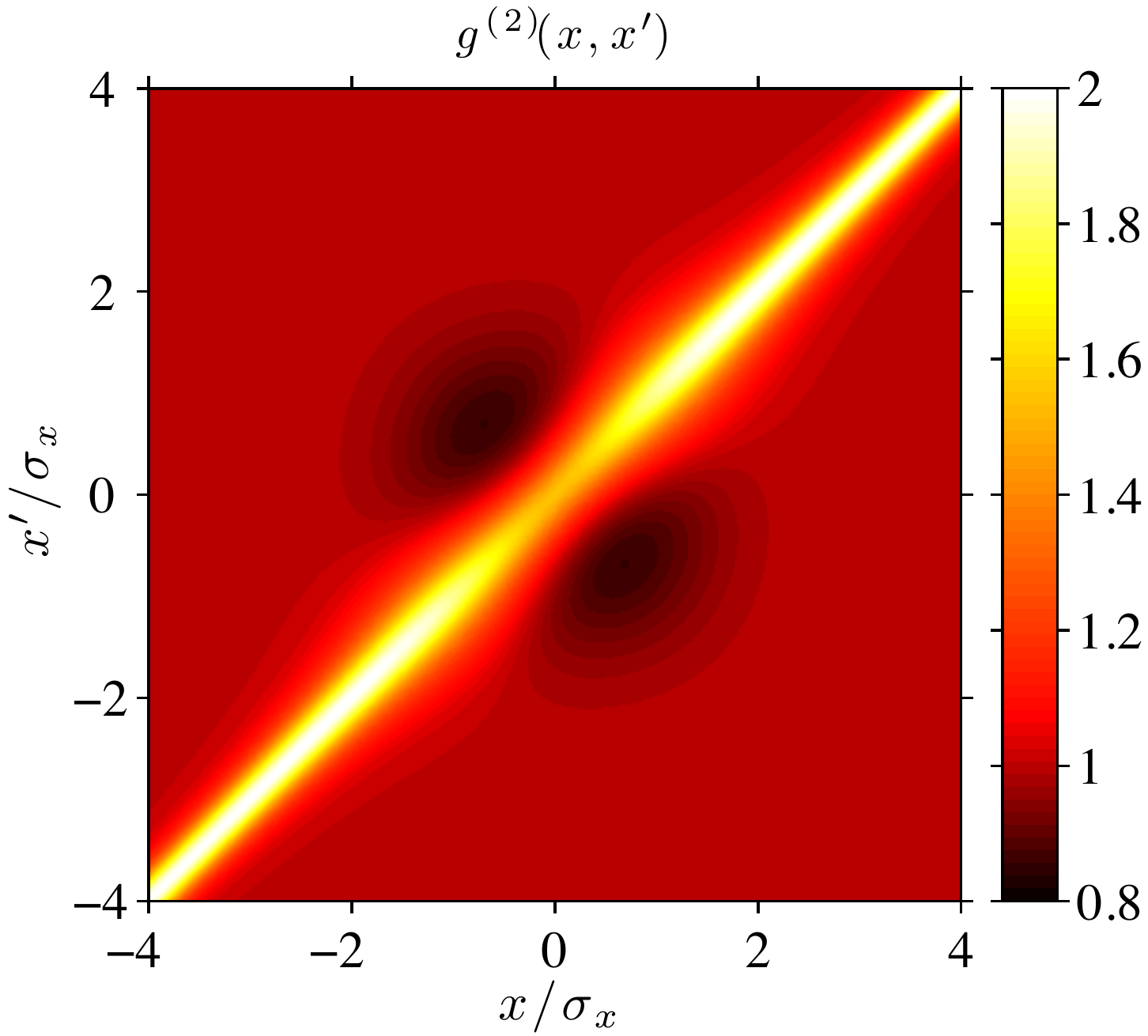}
    \caption{(Color online) Structure of the two-body coherence function $g^{(2)}(x,x^{\prime})\equiv g^{(2)}(x\hat{\mathbf{x}},x'\hat{\mathbf{x}})$ on the long axis of a highly anisotropic trap.  Parameters are $N=100$ atoms, $\omega_\perp/\omega_x=127$, and $T/T_C^0=0.2$ (cf. Fig.~\ref{fig:3}).} 
    \label{fig:5} 
\end{figure}

In Fig.~\ref{fig:5} we plot the full matrix structure $g^{(2)}(x,x^{\prime})\equiv g^{(2)}(x\hat{\mathbf{x}},x'\hat{\mathbf{x}})$ of the coherence function evaluated on the long axis of the anisotropic trapping potential, for the case $N=100$.  Along the diagonal $x'=x$ we observe the suppression of $g^{(2)}(x,x)$ from the HBT value of $2$ (white) at large $|x|$ down to $g^{(2)}(0,0)\approx1.5$ at the origin, due to the presence of the condensate.   The anticorrelation lobes where $g^{(2)}(x,x')<1$ (dark shades of red/gray) are clearly visible, and are centered on the anti-diagonal $x'=-x$ at $|x|\approx \sigma_x$, as expected (see Appendix~\ref{append:physical_origin}).  We note, however, that some (comparatively small) anticorrelation can also be seen at, e.g., $(x,x')=(0,\sigma_x)$, which is evidently due to the intrinsic amplitude anticorrelation of \emph{even} (excited) Hermite modes, illustrating that it is not solely the first Hermite (dipole) mode which contributes to the density anticorrelation feature. 

Finally in this section, we note that we have applied our methodology to the case of $N=100$~atoms in a purely 1D trap, and compared the results for $\min\{g^{(2)}(x,-x)\}$ (which agree very closely with the plotted case of $\omega_\perp/\omega_x=127$) with the correlations of this system estimated using Monte-Carlo sampling of the canonical ensemble in the coherent-state representation~\cite{Heller09}.  These numerical results show qualitative agreement with our semi-analytic calculations, and in fact suggest that the maximal suppression of $g^{(2)}(x,-x)$ is $\sim 20\%$ larger at its peak (as a function of $T$), and that the anticorrelation subsides somewhat more slowly with increasing temperature, due to higher-order contributions to $g^{(2)}(x,x')$ in the canonical ensemble (see Appendix~\ref{append:full_g2}).

\section{Experimental considerations}\label{sec:Relation-to-practice}

The density anticorrelation we discuss in this article is an intrinsically in-trap effect and could therefore be detected, in principle, via an \emph{in situ} measurement of $g^{(2)}(\mathbf{r},\mathbf{r^{\prime}})$ (as proposed, e.g., in Ref.~\cite{Sykes:08}).  Moreover, as the momentum-space correlations of the trapped ideal gas can be obtained (up to a normalization factor) from its position-space correlations by the simple substitution $r_{i}=(\frac{\hbar}{m\omega_{i}})k_{i}$~\cite{Landau-Lifshitz-StatPhys}, where $r_i$ and $k_{i}$ are the $i$th components of the position and momentum vectors $\mathbf{r}$ and $\mathbf{k}$, respectively, the effect can equally well be regarded as an in-trap momentum-space density anticorrelation.   As is well known, position correlations measured in far field after releasing a trapped gas from its confinement reflect (in the limit that interactions can be neglected during the expansion) the corresponding \emph{in situ} momentum correlations of the original trapped sample~\cite{Altman04}.  In fact, in the harmonically trapped non-interacting systems we have considered here, the position-space correlations of the gas are, at all times during the expansion, given by a simple rescaling of the in-trap position-space correlations~\cite{Gomes:06}.  As such, the density anticorrelations we predict for the ideal gas should be observable in near-field expansion as well.  This particularly simple behavior is of course peculiar to the case of an ideal gas, so we now briefly consider the prospects for realizing the ideal-gas limit experimentally, focusing on the quasi-1D regime (Sec.~\ref{subsec:quasi-1D}), in which the anticorrelation dip is most pronounced.   

In a weakly interacting Bose gas in a strongly anisotropic (elongated) trapping geometry, the finite-size condensation behavior~\cite{Ketterle:96} comes into competition with an interaction-driven crossover to a quasicondensate regime, characterized by suppressed density fluctuations but lacking significant nonlocal phase coherence, with a characteristic crossover temperature~\cite{Bouchoule:07}
\begin{equation}
    T_{\mathrm{co}}^{\mathrm{(1D)}}\approx\frac{3N\hbar\omega_{x}}{k_{B}\ln(N\hbar^{3}\omega_{x}/mg^{2})},
\end{equation}
where $g\approx2\hbar\omega_{\perp}a$ is the effective 1D coupling constant~\cite{Olshanii:98} with $a$ the 3D $s$-wave scattering length.  To access the effectively ideal-gas regime, the system parameters must be such that $T_{\mathrm{co}}^{\mathrm{(1D)}} \ll T_C^{\mathrm{(1D)}}$, in which case the system will be well described by the ideal-gas treatment of this article in the temperature range $T_{\mathrm{co}}^{\mathrm{(1D)}} < T < T_C^{\mathrm{(1D)}}$.  As shown in Ref.~\cite{Bouchoule:07}, to access this regime we require $a\lesssim 0.01\;\mathrm{nm}$, which is far smaller than the natural scattering lengths of the alkali-metal atoms; however, this extremely weak scattering could be realized by exploiting Feshbach-resonance techniques~\cite{Chin10}. 

The observation that density anticorrelations can arise even in an ideal Bose gas is particularly relevant in light of recent theoretical and experimental developments concerning the observation of density correlations in time-of-flight, following the release of a low-dimensional (quasicondensed) gas from its confinement~\cite{Imambekov-Petrov:09,Manz:10} (see also Ref.~\cite{Choi12}).  The central observation of these works is that the (axial) coherence $g^{(2)}(x,x')$ of the expanding gas develops ``ripples'' at an intermediate stage of the expansion, which correspond directly to the \emph{in situ} phase fluctuations of the original sample.  As the expansion continues, $g^{(2)}(x,x')$ approaches its far-field limit, which reflects the in-trap momentum correlations of the field.  In Ref.~\cite{Imambekov-Petrov:09}, these momentum correlations are calculated within a local-density approximation, and the far-field limit of the position-space coherence function is therefore found to decay monotonically with $|x-x'|$.  

Although the ideal-gas results presented in this article are not directly applicable to interacting systems, they demonstrate the potentially significant influence of finite-size and inhomogeneity effects on experimentally measured correlations.  Indeed, although density fluctuations are strongly suppressed in quasi-1D interacting systems such as that considered in Ref.~\cite{Manz:10}, phase fluctuations remain significant in such regimes.  In a harmonically trapped quasicondensate, these phase fluctuations exhibit a specific spatial structure resulting from the inhomogeneity of the quasicondensate density profile~\cite{Petrov:00,Petrov:01}.  Accordingly, we might expect such systems to exhibit in-trap momentum-space density anticorrelations~\cite{interacting_case}, which should manifest in position-space measurements made in far field, and may potentially yield corrections to the time development of correlations during free expansion~\cite{Imambekov-Petrov:09}.

More recent experiments~\cite{Perrin-HBT:12} have considered the correlations of interacting atomic bosons in both the strongly anisotropic quasicondensate regime, and in weakly prolate traps in which the physics of the gas is expected to correspond more closely to the comparatively transparent case of strictly three-dimensional Bose-Einstein condensation.  In particular, anticorrelation dips similar to those discussed in this article were observed in time-of-flight expansion of a gas released from $1:4.75$ aspect-ratio confinement, at temperatures $T\sim 0.9T_C$.  Although further analysis would be required to disentangle the effects of interactions and finite expansion times on the measured correlations, it nevertheless seems natural to associate these experimental results with the fundamental physical mechanism discussed in this article: amplitude anticorrelations of excited modes yield density anticorrelations upon interfering with the condensate.   In particular, at temperatures close to $T_C$, we expect that the effects of interactions are relatively subdued, serving at leading order to simply restructure the single-particle excitations of the trap into Hartree--Fock-like modes~\cite{Goldman81}, which should indeed exhibit the same fundamental amplitude anticorrelations that underlie the density anticorrelations we have considered here.

\section{Summary}

In summary, we have predicted the existence of two-body anticorrelations in an ideal Bose gas with a fixed, finite total atom number in harmonic confinement.  The corresponding dip below the uncorrelated level of unity in the second-order coherence function of the Bose field reaches experimentally resolvable magnitudes in highly anisotropic traps containing small atom numbers ($N \lesssim 1000$).  The ideal-gas regime in which the effect is pronounced could be reached in contemporary atomic physics experiments by exploiting Feshbach resonance techniques.  Moreover, the fundamental physical mechanism underlying the anticorrelation should also be relevant to experiments on interacting Bose systems.

\begin{acknowledgments}
T.M.W. and K.V.K. acknowledge support by the ARC Discovery Project grant DP110101047.  A.P. acknowledges support by the ANR project No. 11-PDOC-021-01.  J.S. acknowledges support by the Austrian Science Foundation (FWF) and the Wittgenstein Prize. 
\end{acknowledgments}

\appendix

\section{General form of the second-order coherence function}\label{append:full_g2} 

The second-order coherence function is given in general by~\cite{Nar-Glauber:99}
\begin{align}
    G^{(2)}(\mathbf{r},\mathbf{r}') &= \langle \hat{\Psi}^\dagger(\mathbf{r})\hat{\Psi}^\dagger(\mathbf{r}')\hat{\Psi}(\mathbf{r}')\hat{\Psi}(\mathbf{r})\rangle  \\
    &= \sum_{mnpq} \zeta_m^*(\mathbf{r})\zeta_n^*(\mathbf{r}')\zeta_p(\mathbf{r}')\zeta_q(\mathbf{r}) \langle \hat{a}_m^\dagger \hat{a}_n^\dagger \hat{a}_p \hat{a}_q\rangle, \nonumber  
\end{align}
where $\langle\cdots\rangle\equiv \mathrm{Tr}\{\,\cdots\hat{\rho}\}$ denotes a trace taken with respect to the appropriate thermodynamic density matrix.  In the grand-canonical ensemble the density matrix is of the Gaussian form $\hat{\rho}_G = Z_G^{-1} e^{-(\hat{H}-\mu\hat{N})/kT}$, with $\hat{H}=\sum_n E_n \hat{a}^\dagger_n\hat{a}_n$ and $\hat{N}=\sum_n \hat{a}^\dagger_n\hat{a}_n$, and where $Z_G$ is the grand-canonical partition function, whereas the density matrix appropriate to the canonical ensemble can be written as
\begin{align}\label{eq:rho_canon}
    \hat{\rho}_C = Z_C^{-1} e^{-\hat{H}/kT} \hat{P}_N,
\end{align}
where the projector
\begin{align}
   \hat{P}_N = \sum_{\sum N_m = N} \left| N_0, N_1, \cdots \right\rangle\left\langle N_0, N_1, \cdots \right|,
\end{align}
here expressed in terms of Fock states $| N_0, N_1, \cdots \rangle$ over the basis modes $\zeta_n(\mathbf{r})$, effects the constraint to a fixed total number of particles $N$, and $Z_C$ is the canonical partition function.  As $[\hat{\rho}_G,\hat{N}]=[\hat{\rho}_C,\hat{N}]=0$, in both the canonical and grand-canonical ensembles all so-called anomalous averages ($\langle \hat{a}_m\rangle$, $\langle \hat{a}_m \hat{a}_n\rangle$, $\langle \hat{a}_m^\dagger \hat{a}_n \hat{a}_p \rangle$, etc.) vanish, and all two-body correlations of the field are therefore completely specified by the one-body correlations $\langle \hat{a}_m^\dagger \hat{a}_n\rangle = \delta_{mn} N_m$ and the cumulants 
\begin{align}
    C_{mnpq} &\equiv \langle \hat{a}_m^\dagger \hat{a}_n^\dagger \hat{a}_p \hat{a}_q \rangle  \\
    &\qquad- \langle \hat{a}_m^\dagger \hat{a}_p\rangle \langle \hat{a}_n^\dagger \hat{a}_q\rangle - \langle \hat{a}_m^\dagger \hat{a}_q\rangle \langle \hat{a}_n^\dagger \hat{a}_p\rangle \nonumber \\
    &= \langle \hat{a}_m^\dagger \hat{a}_n^\dagger \hat{a}_p \hat{a}_q \rangle - (\delta_{mp}\delta_{nq}+\delta_{mq}\delta_{np})N_mN_n, \nonumber  
\end{align}
which quantify the deviation of the field fluctuation statistics from the Gaussian limit.  In the grand-canonical ensemble, these cumulants vanish identically, so the second-order coherence function is specified completely by the occupation numbers $N_m$, and can be written~\cite{Nar-Glauber:99} 
\begin{align}
    \left[G^{(2)}(\mathbf{r},\mathbf{r}')\right]_G &= \langle \hat{\Psi}^\dagger(\mathbf{r})\hat{\Psi}(\mathbf{r})\rangle \langle \hat{\Psi}^\dagger(\mathbf{r}')\hat{\Psi}(\mathbf{r}')\rangle \nonumber \\
    &\qquad + \langle \hat{\Psi}^\dagger(\mathbf{r})\hat{\Psi}(\mathbf{r}')\rangle \langle\hat{\Psi}^\dagger(\mathbf{r}')\hat{\Psi}(\mathbf{r})\rangle. 
\end{align} 
In the canonical ensemble we have, by the form of the density operator $\hat{\rho}_C$ [Eq.~(\ref{eq:rho_canon})], $C_{mnpq}=0$ unless $(m,n)=(p,q)$ or $(m,n)=(q,p)$.  Thus in the case $m=n$, a nonzero cumulant is found only when $p=q=m$.  Therefore, the second-order coherence function in the canonical ensemble $[G^{(2)}(\mathbf{r},\mathbf{r}')]_C = [G^{(2)}(\mathbf{r},\mathbf{r}')]_G + \Delta G^{(2)}(\mathbf{r},\mathbf{r}')$, where  
\begin{align}\label{eq:deltaG2}
    \Delta G^{(2)}(\mathbf{r},\mathbf{r}') &= \sum_{mnpq} \zeta_m^*(\mathbf{r})\zeta_n^*(\mathbf{r}')\zeta_p(\mathbf{r}')\zeta_q(\mathbf{r}) C_{mnpq}  \\
    &= \sum_m |\zeta_m(\mathbf{r})|^2|\zeta_m(\mathbf{r}')|^2 C_{mmmm} \nonumber \\
    &\qquad + \sum_m\sum_{n\neq m} \Big[ |\zeta_m(\mathbf{r})|^2 |\zeta_n(\mathbf{r}')|^2 C_{mnnm} \nonumber \\
    &\qquad\qquad + \zeta_m^*(\mathbf{r})\zeta_n^*(\mathbf{r}')\zeta_m(\mathbf{r}')\zeta_n(\mathbf{r}) C_{mnmn}\Big]. \nonumber
\end{align}
To proceed, we separate off terms in Eq.~(\ref{eq:deltaG2}) that involve the condensate mode $\zeta_0(\mathbf{r})$, and use the symmetries of the cumulants $C_{mnpq}$ and the relations 
\begin{align}
    \langle \Delta N_m \Delta N_n \rangle &\equiv \left\langle \left(N_m - \langle N_m \rangle\right)\left(N_n - \langle N_n\rangle\right)\right\rangle \nonumber \\
    &= \left\{
    \begin{array}{ll}
        \!C_{mmmm} + N_m(N_m+1) & \quad m=n \\ 
        \!C_{mnmn} & \quad m\neq n,
    \end{array} \right.
\end{align}
to obtain (after some algebra) the exact form
\begin{widetext}
\begin{align}\label{eq:full_G2}
    \left[G^{(2)}(\mathbf{r},\mathbf{r}')\right]_C &= \langle \hat{\Psi}^\dagger(\mathbf{r})\hat{\Psi}(\mathbf{r})\rangle \langle \hat{\Psi}^\dagger(\mathbf{r}')\hat{\Psi}(\mathbf{r}')\rangle + \langle \hat{\Psi}^\dagger(\mathbf{r})\hat{\Psi}(\mathbf{r}')\rangle \langle\hat{\Psi}^\dagger(\mathbf{r}')\hat{\Psi}(\mathbf{r})\rangle + \left[ \langle (\Delta N_0)^2 \rangle - N_0(N_0+1) \right]|\zeta_0(\mathbf{r})|^2|\zeta_0(\mathbf{r}')|^2  \nonumber \\ 
    &\qquad + \sideset{}{'}\sum_m \langle \Delta N_0 \Delta N_m \rangle \left| \zeta_0(\mathbf{r})\zeta_m(\mathbf{r}') + \zeta_0(\mathbf{r}')\zeta_m(\mathbf{r}) \right|^2 \nonumber \\
    &\qquad+\sideset{}{'}\sum_m \left[ \langle (\Delta N_m)^2 \rangle - N_m(N_m+1)\right]|\zeta_m(\mathbf{r})|^2|\zeta_m(\mathbf{r}')|^2 \nonumber \\ 
    &\qquad+\sideset{}{'}\sum_m\sideset{}{'}\sum_{n\neq m} \langle \Delta N_m \Delta N_n\rangle\left[ |\zeta_m(\mathbf{r})|^2|\zeta_n(\mathbf{r}')|^2 + \zeta_m^*(\mathbf{r})\zeta_n^*(\mathbf{r}')\zeta_m(\mathbf{r}')\zeta_n(\mathbf{r}) \right],
\end{align}
\end{widetext}
where $\sum_m'$ denotes a sum which excludes the condensate mode ($m=0$).  Terms on the first line of Eq.~(\ref{eq:full_G2}) constitute the approximate form for the coherence function that we use in the main text of this article, from which the normalized form [Eq.~(\ref{eq:g2-IBG})] is obtained as $g^{(2)}(\mathbf{r},\mathbf{r}') \equiv [G^{(2)}(\mathbf{r},\mathbf{r}')]_C/[G^{(1)}(\mathbf{r},\mathbf{r})G^{(1)}(\mathbf{r}',\mathbf{r}')]$, where $G^{(1)}(\mathbf{r},\mathbf{r}')\equiv\langle\hat{\Psi}^\dagger(\mathbf{r})\hat{\Psi}(\mathbf{r}')\rangle$~\cite{Nar-Glauber:99}.  This approximate form for $g^{(2)}(\mathbf{r},\mathbf{r}')$ generalizes those given in Ref.~\cite{Nar-Glauber:99} by allowing for fluctuation statistics of the condensate occupation between the limiting cases of a purely thermal or purely coherent state.

The terms on the second line of Eq.~(\ref{eq:full_G2}) correspond to the effects of explicit correlations between the fluctuations of the condensate occupation and those of the excited-mode populations.  Politzer~\cite{Politzer:96} has shown that these correlations are negative, i.e., that fluctuations of the excited-mode populations are anticorrelated with those of the condensate occupation.  The anticorrelations between mode occupations are  most pronounced near the Bose-condensation transition, and the largest such anticorrelation is that between the condensate and the (degenerate) first excited state, for which $-\langle \Delta N_0 \Delta N_1 \rangle \lesssim 0.1 N_0 N_1$ for $N=100$ (or more) atoms in an isotropic trap~\cite{Politzer:96}.  The spatial term multiplying $\langle \Delta N_0 \Delta N_m\rangle$ in Eq.~(\ref{eq:full_G2}) is manifestly non-negative, and these anticorrelations therefore yield a small additional \emph{suppression} of the second-order coherence function relative to Eq.~(\ref{eq:g2-IBG}); i.e., they do not subtract from the anticorrelation dip we consider in this article, and can thus be safely neglected.

The terms on the third line of Eq.~(\ref{eq:full_G2}) correspond to the deviations of the number fluctuations of each of the excited modes from their grand-canonical levels.  In fact, in the canonical ensemble these fluctuations are somewhat suppressed below the grand-canonical predictions --- an effect easily understood in terms of the effective-interaction model of the canonical-ensemble constraint discussed by Wilkens and Weiss~\cite{Wilkens1} --- and these terms therefore do not subtract from the anticorrelation dip either.  The final line of Eq.~(\ref{eq:full_G2}) corresponds to the correlations between excited-mode population fluctuations.  These correlations are again negative, and although the spatial term multiplying $\langle \Delta N_m \Delta N_n\rangle$ is not strictly positive, these anticorrelations are weak (smaller in magnitude than $\langle \Delta N_0 \Delta N_1\rangle$~\cite{Politzer:96}), and yield only a small quantitative correction to $g^{(2)}(\mathbf{r},\mathbf{r}')$.

\section{Physical origin of the effect}\label{append:physical_origin} 

\looseness+1 The physical origin of the anticorrelation dip in the second-order coherence function is the intrinsic \emph{amplitude} anticorrelation of the excited energy eigenstates of the trap.  The underlying mechanism can be illustrated using a simple two-mode model, consisting of the first two energy eigenmodes of a 1D harmonic trap; i.e., the ground mode $\zeta_0(x) = (\sqrt{\pi}\sigma_x)^{-1/2}\exp(-x^2/2\sigma_x^2)$ and the ``dipole'' mode $\zeta_1(x) = (2/\sqrt{\pi}\sigma_x)^{1/2} (x/\sigma_x) \exp(-x^2/2\sigma_x^2)$.  We make minimal assumptions as to the correlations of the two modes: we assume that there are no amplitude or number correlations between the two modes ($\langle \hat{a}_0^\dagger \hat{a}_1\rangle=0$ and $\langle \hat{a}_0^\dagger \hat{a}_0 \hat{a}_1^\dagger \hat{a}_1 \rangle = \langle\hat{a}_0^\dagger \hat{a}_0\rangle \langle\hat{a}_1^\dagger \hat{a}_1\rangle$, respectively), and that the excited mode is incoherent, $\langle \hat{a}_1^\dagger\hat{a}_1^\dagger\hat{a}_1\hat{a}_1\rangle = 2\langle \hat{a}_1^\dagger \hat{a}_1\rangle^2\equiv 2N_1^2$, while allowing for partial coherence of the condensate mode, $\langle \hat{a}_0^\dagger \hat{a}_0^\dagger \hat{a}_0 \hat{a}_0 \rangle = g_0^{(2)} \langle \hat{a}_0^\dagger \hat{a}_0\rangle^2\equiv g_0^{(2)}N_0^2$.  With these basic assumptions, one finds that the second-order coherence function is given (exactly) by Eq.~(\ref{eq:g2-IBG}), where $G^{(1)}(x,x')=\sum_{i=0}^1N_i\zeta_i(x)\zeta_i(x')$, $\rho(x)=G^{(1)}(x,x)$, and $\rho_0(x)=N_0\zeta_0^2(x)$.  We consider the second-order coherence between the turning points $\pm \sigma_x$ of the dipole mode (between which the amplitude anticorrelation is largest): we have $\zeta_0(\pm\sigma_x)=(e^{-1}/\sqrt{\pi}\sigma_x)^{1/2}$, and $\zeta_1(\sigma_x)=-\zeta_1(-\sigma_x)=(2e^{-1}/\sqrt{\pi}\sigma_x)^{1/2}$, and thus
\begin{align}
    g^{(2)}(\sigma_x,-\sigma_x) = 1+ \frac{(N_0-2N_1)^2 - (2-g_0^{(2)})N_0^2}{(N_0+2N_1)^2}.
\end{align}
Clearly, in the case that the condensate mode is incoherent ($g_0^{(2)}=2$), $g^{(2)}(\sigma_x,-\sigma_x)\geq 1$.  To obtain $g^{(2)}(\sigma_x,-\sigma_x)<1$ we in fact require $g_0^{(2)} < 2 - (1-2N_1/N_0)^2$.  Thus partial coherence of the condensate is required in order for the dip to arise, which explains why the anticorrelation is only visible for $T\lesssim T_C$, and is obscured in a grand-canonical treatment.  Moreover, as $N_1$ becomes small compared to $N_0$, the value of $g_0^{(2)}$ required to observe the dip becomes smaller, consistent with the observation that the anticorrelation effect weakens as $T\to0$.  Considering the case that the condensate is perfectly coherent $g^{(2)}_0=1$, in the limit $N_1\ll N_0$ we obtain $g^{(2)}(\sigma_x,-\sigma_x)\approx 1 - 4N_0N_1/(N_0+2N_1)^2 \approx 1 - 4N_0N_1/N^2 \approx 1-4N_1/N$ (where $N=\sum_i N_i$) --- which gives a useful characterization of the dependence of the magnitude of the dip on the occupations of the two modes.  This simple model obviously generalizes to the case of the gas in thermodynamic equilibrium, in which higher trap modes also produce density anticorrelations upon interfering with the condensate --- though the dipole modes are expected to yield the dominant contribution to the effect, due to the relative magnitude of both their antinodes and their occupations at equilibrium.   Finally, we note that the anticorrelation effect can easily be understood in terms of the classical picture of HBT interferometry of optical fields~\cite{Baym98}:  Consider the field $\alpha(x)$ radiated by an incoherent source $a$ ($\langle |\alpha|^4\rangle = 2\langle |\alpha|^2\rangle^2$), at two points $x_1$ and $x_2$ between which $\alpha(x)$ exhibits a $\pi$ phase difference [$\alpha(x_2)=-\alpha(x_1)$].  In the presence of a second field $\beta(x)$, uncorrelated with $\alpha(x)$, that exhibits zero relative phase between $x_1$ and $x_2$, the measured intensities at the two points will exhibit an anticorrelation $\langle I_1 I_2 \rangle < \langle I_1 \rangle \langle I_2 \rangle$, provided that the field $\beta$ exhibits sufficient coherence $g^{(2)}_\beta \equiv \langle |\beta|^4\rangle/\langle |\beta|^2\rangle^2 < 2 - (\langle|\alpha|^2\rangle-\langle|\beta|^2\rangle)^2/\langle|\beta|^2\rangle^2$ to overcome the HBT bunching effect.  The only difference in the present case of a trapped Bose gas is that no propagation is necessary to observe the effect, as the phase differences between spatial points arise naturally from the structure of harmonic-oscillator energy eigenmodes.

\bibliographystyle{prsty}

\begin{thebibliography}{10}
\bibitem{NoteB}
{Note that although we use the exact canonical-ensemble value of $N_0$ in
  calculating $g_0^{(2)}$, the approximate grand-canonical value for the
  condensate occupation is implicit in the other terms appearing in Eq.~(2)
  [see Eqs.~(3)--(6)]. An improved approximation can be obtained by subtracting
  off the resulting discrepancy from $\rho(\mathbf{r})$, $\rho_0(\mathbf{r})$,
  and $G^{(1)}(\mathbf{r},\mathbf{r}')$, and adjusting $\mu$ accordingly, but
  we have found that this yields only a small quantitative correction to the
  anticorrelation dip. Thus we present here the more compact and
  straightforward analysis in which the grand-canonical value for $N_0$ is used
  in all quantities except for $g_0^{(2)}$.}
\vspace{2\baselineskip}
\end{thebibliography}

\end{document}